\documentclass[aps, prl, amsmath,amssymb, reprint, showpacs,superscriptaddress,groupedaddress, nofootinbib ]{revtex4-1}

\usepackage{graphicx}
\usepackage{dcolumn}
\usepackage{bm}
\usepackage{amsmath}
\usepackage{color}

\usepackage{dcolumn}  
\usepackage{amsmath}
\usepackage{multirow}
\usepackage{graphicx}   
\usepackage{subfigure}
\usepackage{comment}
\usepackage{color}
\usepackage[colorlinks,bookmarks=false,citecolor=blue,linkcolor=red,urlcolor=blue]{hyperref}
\usepackage{nicefrac}
\usepackage{soul}
\usepackage{slashbox}
\usepackage{bm,array}
\usepackage{tikz}
\usepackage{pifont}
\usepackage{gensymb}
\newcommand{\cmk}{\textcolor{green}{\ding{51}}}
\newcommand{\xmk}{\textcolor{red}{\ding{55}}}

\newcolumntype{C}{>{\centering\arraybackslash}p{1em}}

\begin{document}
	
	\title{Double Perovskites overtaking the single perovskites : A set of new solar harvesting materials with much higher stability and efficiency}
	\author{Jiban Kangsabanik,$^1$ Vipinraj Sugathan,$^2$ Anuradha Yadav,$^1$ Aswani Yella,$^2$ Aftab Alam$^1$}
	\email{aftab@iitb.ac.in}	
	\affiliation{$^1$Department of Physics, Indian Institute of Technology Bombay, Mumbai 400076, India}
	\affiliation{$^2$Department of Metallurgical Engineering \& Materials Science, Indian Institute of Technology Bombay, Mumbai 400076, India}
	
	\begin{abstract}
		Solar energy plays an important role in substituting the ever declining source of fossil fuel energy. Finding novel materials for solar cell applications is an integral part of photovoltaic research. Hybrid Lead halide perovskites are one of, if not the most, well sought material in the photovoltaic research community. Its unique intrinsic properties, flexible synthesis techniques, and device fabrication architecture made the community highly buoyant over the past few years. Yet, there are two fundamental issues which still remain a concern  i.e. stability in external environment and toxicity due to Pb. This led to a search for alternative materials. More recently, double perovskite (A$_2$BB$^{'}$X$_6$ (X=Cl,Br,I)) materials have emerged as a promising choice. Few experimental synthesis and high throughput computational studies have been carried out to check for promising candidates of this class. The main outcome from these studies, however, can essentially be summarized into two categories, (i) either they have indirect band gap or (ii) direct but large optical band gap which are not suitable for solar device. Here we propose a large set of stable double perovskite materials, Cs$_2$BB$^{'}$X$_6$ (X=Cl,Br,I), which show indirect to direct band gap transition via small Pb$^{+2}$ doping at both B and B$^{'}$ sites. This is done by careful band (orbital) engineering using first principles calculations. This kind of doping has helped to change the topology of band structure triggering the indirect to direct transition which are optically allowed. It also reduces the band gap significantly, bringing it well in the visible region.  We also simulated the optical absorption spectra of these systems and found comparable/higher absorption coefficient and solar efficiency with respect to the state of the art photovoltaic absorber material CH$_3$NH$_3$PbI$_3$. A number of materials  Cs$_2$(B$_{0.75}$Pb$_{0.25}$)(B$^{'}$$_{0.75}$Pb$_{0.25}$)X$_6$ (for various combinations of B, B$^{'}$ \& X) are found to be promising, some with better stability and solar efficiency than CH$_3$NH$_3$PbI$_3$, but with much less toxicity. Experimental characterization of one of the material, Cs$_2$(Ag$_{0.75}$Pb$_{0.25}$)(Bi$_{0.75}$Pb$_{0.25}$)Br$_6$ is carried out. Measured properties such as band gap and chemical stability agrees fairly well with the  theoretical predictions. This material is shown to be even more stable than  CH$_3$NH$_3$PbI$_3$, both under the sufficient humidity ($\sim$55\%) and temperature (T=338 K), and hence has potential to become better candidate than the state of the art material. 
		
	\end{abstract}

	\maketitle
	
	\section*{I. Introduction }
	
	As most of the fossil fuel energy sources are declining towards an end, quest for natural, renewable energy sources is becoming an area of utmost importance. At this moment, solar energy plays an integral role here, as such photovoltaics is one of the most important field of research. Finding novel materials for solar cell
	applications, is an elemental part of it. Hybrid Lead halide perovskites are one of, if not the most well sought material in the photovoltaic research community in recent past. Starting from 3.8\% efficiency at its beginning in 2009,\cite{kojima2009organometal} it has now reached a reported efficiency of 22.1\% within only 7 years.\cite{qian2016comprehensive} It's unique intrinsic properties as well as very flexible synthesis technique and device fabrication architecture made the community highly buoyant over the past few years. Yet, there remain two fundamental issues i.e. stability in external environment  and toxicity due to Pb. Various studies, both at experimental and computational level have been carried out to solve these. From the device fabrication perspective, improving the quality of the film using different conducting layers, encapsulation, etc., helped to increase the stability to some extent. Also changing the organic cation, mixing different halides helped to slightly improve the stability, but compromised the efficiency and other properties in many cases.\cite{yin2015halide} Next, to remove the toxicity many people have tried to find suitable substitution of Pb at B site of ABX$_3$ structure.\cite{frolova2016exploring} But that either resulted in higher band gap value (in case of alkaline earth metals)\cite{kumar2016crystal} which is unsuitable as an absorber material, or they turned out to be even more unstable (for Sn and Ge substitutions).\cite{noel2014lead, yokoyama2016overcoming} This led to a search for alternative materials to achieve both the necessary criteria. In 2016, Volonakis et al. suggested heterovalent substitution of Pb with one +3 and one +1 element resulting in double perovskite structure.\cite{volonakis2016lead} Slavney et al. first synthesized and reported Cs$_2$AgBiBr$_6$ to have long carrier lifetime and band gap in visible range.\cite{slavney2016bismuth} But the material shows indirect nature of band gap with poor absorption properties. In addition, it is prone to high non-radiative recombination loss, making it less suitable for solar harvesting purpose. Next, Volonakis et al. reported Cs$_2$InAgCl$_6$, which although has a direct band gap,\cite{volonakis2017cs2inagcl6} but the gap value was on the high violet region. Since then, various computational studies were carried out exploring numerous combinations of halide double perovskites.\cite{jain2017high, chakraborty2017rational, zhao2017design} The main outcome from these studies are essentially similar to the two  specific cases described above i.e. either the materials have indirect band gap or direct but large optical band gap. Recently, Meng et al. in their computational study reported that for most of the direct band gap halide double perovskites, the transition from valence band maximum (VBM) to conduction band minimum (CBM) is optically forbidden. This happens due to the inversion symmetry present in this structure, resulting in VBM and CBM to have same parity.\cite{meng2017parity} They only found Cs$_2$(In,Tl)$^{+1}$(Sb,Bi)$^{+3}$X$_6$ to be the only class with optically allowed direct band gaps. These compounds are also reported by Zhao et al. in their study, to have superior theoretical efficiency.\cite{zhao2017design} But Indium is prone to remain in In$^{+3}$ oxidation state and hence the above compound is literally impossible to form.\cite{xiao2017intrinsic} From the above discussion one can conclude that, for the double perovskites the next possible avenue should be to find a way to either decrease the optically allowed band gaps or to band engineer an indirect band gap system to achieve a direct band gap. Recently, some studies have been reported to control the nature of the band gap.\cite{tran2017designing,slavney2017defect} 
	
	Here we tried to solve this issue with a different approach. In the perovskites structure ABX$_3$, the inherent inversion symmetry causes the possibility of parity forbidden transitions, which can substantially affect the performance of a solar absorber material. This is precisely the case for most of the double perovskites. In the perovskite structure the BX$_6$ octahedra plays an important role in dictating various symmetry properties. Tilting or twisting the octahedra can affect the symmetry properties in a particular material. Here the relative size of A and B cation (their imbalance to be precise) can cause certain tilting or twisting of the octahedra which may break the inversion symmetry. Here we substituted a bivalent element to both the B$^{+1}$ and B$^{'+3}$ site in a double perovskite structure Cs$_2$BB$^{'}$X$_6$ (X=Cl,Br,I) and study their effect on the crystal structure, electronic and optical properties. Although we have simulated a number of compounds Cs$_2$BB$^{'}$X$_6$ involving various combinations of B, B$^{'}$ \& X (as shown in Table I), let us choose a representative example, Cs$_2$AgBiX$_6$ (X= Cl, Br, I), which has been widely studied in the literature. This particular material has been synthesized (except when X=I), and also shown to be chemically stable.\cite{zhao2017design}. Nevertheless, this compound could not stand out as a solar absorber material because of large value and indirect nature of its band gap. However when we substituted 25\% Pb element in place of both Ag$^{+1}$ and Bi$^{+3}$ sites (to balance charge neutrality), it not only reduces the value of band gap but also makes it direct in nature. In addition the transition at direct band gap is found to be optically allowed, unlike many other compounds of the same class. The drastic effect of Pb-substitution can be seen from Fig.2, which shows the band structure and transition probability (square of dipole transition matrix elements) for Cs$_2$AgBiCl$_6$ and its Pb doped counterpart.

	Then we went on to simulate the optical properties of this system and found the results to be very interesting. `Spectroscopic Limited Maximum Efficiency' (SLME) proposed by Yu et al.\cite{yu2012identification} is a well known parameter to quantify the performance of a photovoltaic absorber. SLME, which takes material specific characteristic properties (band structure, absorption coefficient, etc.) into account while calculating the theoretical efficiency limit, can be seen as an improvement upon the widely known theoretical Shockley-Queisser (S-Q) limit.\cite{shockley1961detailed}(detailed discussion about SLME can be found in supplementary material).\cite{supplement} This new material is found to show high value of absorption coefficients and SLME, comparable to the state of the art hybrid perovskites. Next, we simulated the effect of Pb substitution on a class of possible indirect band gap double perovskites and found a number of promising compounds suitable for solar applications. Few of these compounds turn out to be even more promising than state of the art lead halide perovskites. One such compound is Pb-doped Cs$_2$AgBiBr$_6$ which we prepared in our lab and measured its electronic  properties. The measured structure,  magnitude and direct nature of the band gap confirms our simulated data. Interestingly, this material shows a much higher stability (both against humidity and temperature) than CH$_3$NH$_3$PbI$_3$.

	The rest of the paper is organised as follows,  Details about crystal structure of the concerned compounds and the primary screening is given in Section II. Next, the electronic structure and related discussions are  provided in Section III. We discuss the optical properties and SLME comparison in sec IV. Section V is devoted for details of experimental results for Pb-doped Cs$_2$AgBiBr$_6$. In section VI, we explain the sensitivity of efficiency of these systems to their band gaps along with the conclusion and discussions.  The computational and experimental methodology is  discussed in section VII.

	\begin{table*}[t!]
		\begin{ruledtabular}
			\begin{centering}
				\begin{tabular}{||c|C|C|C|C|C|C|C|C|C|C|C|C||}
					\hline
					\multirow{2}{*}{\backslashbox{B}{B$^{'}$}}
					& \multicolumn{3}{c|}{Sb} & \multicolumn{3}{c|}{Bi} & \multicolumn{3}{c|}{Sc} & \multicolumn{3}{c|}{Y} \\
					\cline{2-13}
					& Cl & Br & I & Cl & Br & I & Cl & Br & I & Cl & Br & I \\
					\hline 
					Cu & \cmk & \xmk & \xmk &\cmk & \xmk & \xmk &\cmk & \xmk & \xmk &\cmk & \xmk & \xmk \\
					\hline
					Ag &\cmk & \xmk & \xmk &\cmk &\cmk &\cmk &\cmk & \xmk & \xmk &\cmk &\cmk &\cmk \\
					\hline
					Au &\cmk & \xmk & \xmk &\cmk &\cmk &\cmk &\cmk & \xmk & \xmk &\cmk &\cmk &\cmk \\
					\hline
				\end{tabular}
			\end{centering}
			\caption{Possible permutations of G-XI elements (Cu, Ag, Au) at B site, and G-XV (Sb, Bi) and G-III (Sc,Y) elements at B$^{'}$site in forming the double perovskite compound Cs$_2$BB$^{'}$X$_6$ (X=Cl,Br,I). \cmk\ means the compound satisfying both the Goldschmidt's \& octahedral tolerance criteria, whereas \xmk\      represents the structure not satisfying either or both of these criteria. }
		\end{ruledtabular}	
	\end{table*}

	\section*{II. Structural details and primary screening}
	
	ABX$_3$ perovskites have cubic structure where all the B atoms are surrounded by six nearest X atoms forming an octahedra, while A atoms are surrounded by twelve nearest X atoms. For the double perovskites A$_2$BB$^{'}$X$_6$, the structure remains similar, with alternate BX$_6$ and B$^{'}$X$_6$ octahedra. A$_2$BB$^{'}$X$_6$ (X= Cl, Br, I) crystallizes in cubic structure with space group Fm-3m (\#225). Here `A' occupies the 8c Wyckoff site, B \&  B$^{'}$ takes the 4a \& 4b Wyckoff sites respectively, while halide element sits at 24e site. For halide double perovskites A \& B can be elements having charge state +1, B$^{'}$ are elements with charge state +3 and X are halides (Cl, Br, I) having charge state -1. In the present study, for the parent compounds we choose A as Cs, B as G-XI elements (Cu, Ag, Au), and B$^{'}$ as G-XV (Sb, Bi) and G-III (Sc,Y) elements. For these compounds to be stable in high symmetry cubic structure, two basic tolerance criteria need to be satisfied. The first one is Goldschmidt's tolerance factor t(=(R$_A$+R$_X$)/$\sqrt2$(R$_B$+R$_X$)), which should be  higher than 0.8 and close to 1. There is another factor called octahedral factor $\mu$(=R$_B$/R$_X$), which should be larger than 0.414 for BX$_6$ octahedra to be stable.\cite{li2008formability} Here R$_A$, R$_B$, and R$_X$ are the ionic radii of the elements sitting at A, (B \& B$^{'}$), and X sites respectively. The first level of screening is done by choosing those parent compounds which satisfy both of these criteria. (shown in Table I). The radius of B and B$^{'}$ elements are taken separately, and the tolerance criteria is checked for both the octahedra individually. Next, we substitute Pb in these structurally stable parent compounds. Pb, having ionic radius of 1.19\ \r{A} satisfies the tolerance criteria for all the halides.\cite{shannon1969effective}

	For the Pb substitution, we took a 40 atom cubic conventional cell for each stable structure consisting of 8 A cations, 4 B cations, 4 B$^{'}$ cations, and 24 X anions. Then one out of each of four B \& B$^{'}$ sites are replaced with a Pb atom. In the conventional cell B element occupies the positions (in fractional co-ordinates) (0, 0, 0), (0.0, 0.5, 0.5), (0.5, 0.0, 0.5), (0.5, 0.5, 0.0) and B$^{'}$ element occupies the positions (0.5, 0.5, 0.5), (0.5, 0.0, 0.0), (0.0, 0.5, 0.5), (0.0, 0.0, 0.5). Here we found two possible inequivalent combinations of substitutions. In the first case, we substitute Pb at B (0, 0, 0) \& B$^{'}$ (0.5, 0.5, 0.5) site. In the second case we put Pb at B (0, 0, 0) \& B$^{'}$ (0.5, 0.0, 0.0) site. These two configurations are shown in Figure 1. All the other configurations are energetically degenerate to one of these structures and hence are equivalent. We will denote the first structure as structure `1' and the second structure as structure `2' from here onwards.

	\begin{figure*}[t!]
		\centering
		\includegraphics[width=0.90\linewidth]{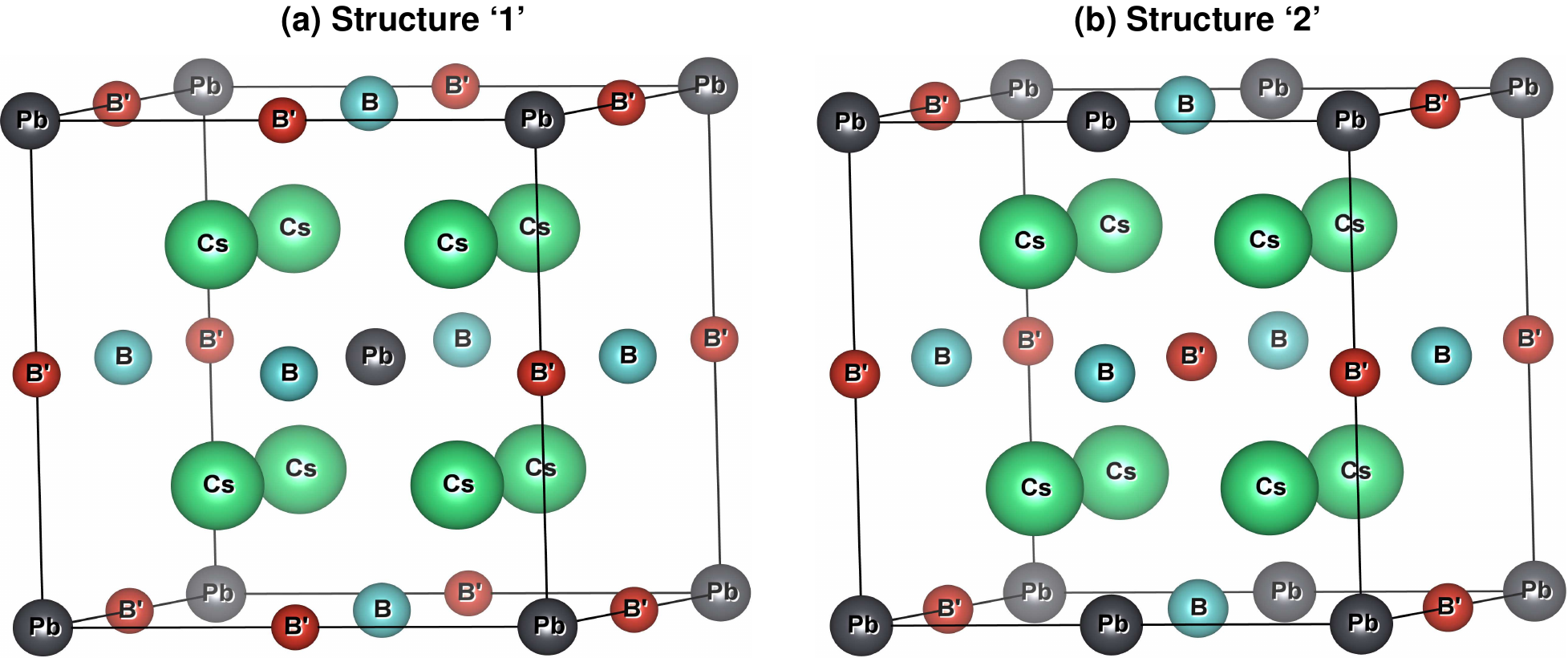}
		\caption{A conventional unit cell for Pb-doped Cs$_2$BB'X$_6$. (a) and (b) are the two possible distinct configurations depending on the positions of Pb substituted at  B, and B$^{'}$ sites. Here Cs, B, B$^{'}$ \& Pb atoms are shown using green, turquoise, red, and grey coloured spheres respectively. Halogen atoms are omitted for convenience. }
		\label{fig:test2}
	\end{figure*}	
	
	First principles calculations were performed by using Density Functional Theory (DFT)\cite{kohn1965self} with Projector Augmented Wave (PAW) basis set\cite{blochl1994projector} as implemented in Vienna Ab-initio Simulation Package (VASP).\cite{kresse1996efficiency,kresse1999ultrasoft} Pseudopotential formalism with Perdew-Burke-Ernzerhof (PBE) exchange correlation functional\cite{perdew1996generalized} is used to do the primary electronic structure calculations. A more accurate estimation of band gap and other optical properties are carried out using hybrid Heyd-Scuseria-Ernzerhof (HSE) exchange correlation functional including the spin orbit coupling (soc). More computational details can be found in the supplementary material.\cite{supplement} Our calculated equilibrium lattice parameters for the well-studied systems Cs$_2$AgBiCl$_6$ (10.94 \r{A}) and Cs$_2$AgBiBr$_6$ (11.46 \r{A}) matches very well with the previously reported experimental data (10.78 \r{A} and 11.27 \r{A}).\cite{mcclure2016cs2agbix6} Small discrepancies in the value can be attributed to overestimation of lattice parameter by GGA exchange-correlation functional, which is well-known. After relaxation, structure `1'  retains its cubic symmetry, but structure `2' is found to elongate slightly along the direction where both the parent B and B$^{'}$ elements are replaced by Pb (here x-direction). This can be due to the fact that Pb (1.19 \r{A}) has higher atomic radius  than all the other B and B$^{'}$ parent elements studied in this work, except for Au (1.37 \r{A}). Even in case of Au at B-site, the average (1.165 \r{A}) for  R$_B$ \& R$_{B^{'}}$ for the biggest atom Bi (0.96 \r{A}) at B$^{'}$ site is lower than the atomic radius of Pb. This (bigger Pb atom) also accounts for the increased volume for the substituted compounds than the parent compounds. A more detailed discussion on structural formation based on bond length and electronegativity, along with the theoretically relaxed lattice parameters are reported in supplementary material  (see section S2).\cite{supplement}

	\begin{figure*}[t!]
		\centering
		\includegraphics[width=0.90\linewidth]{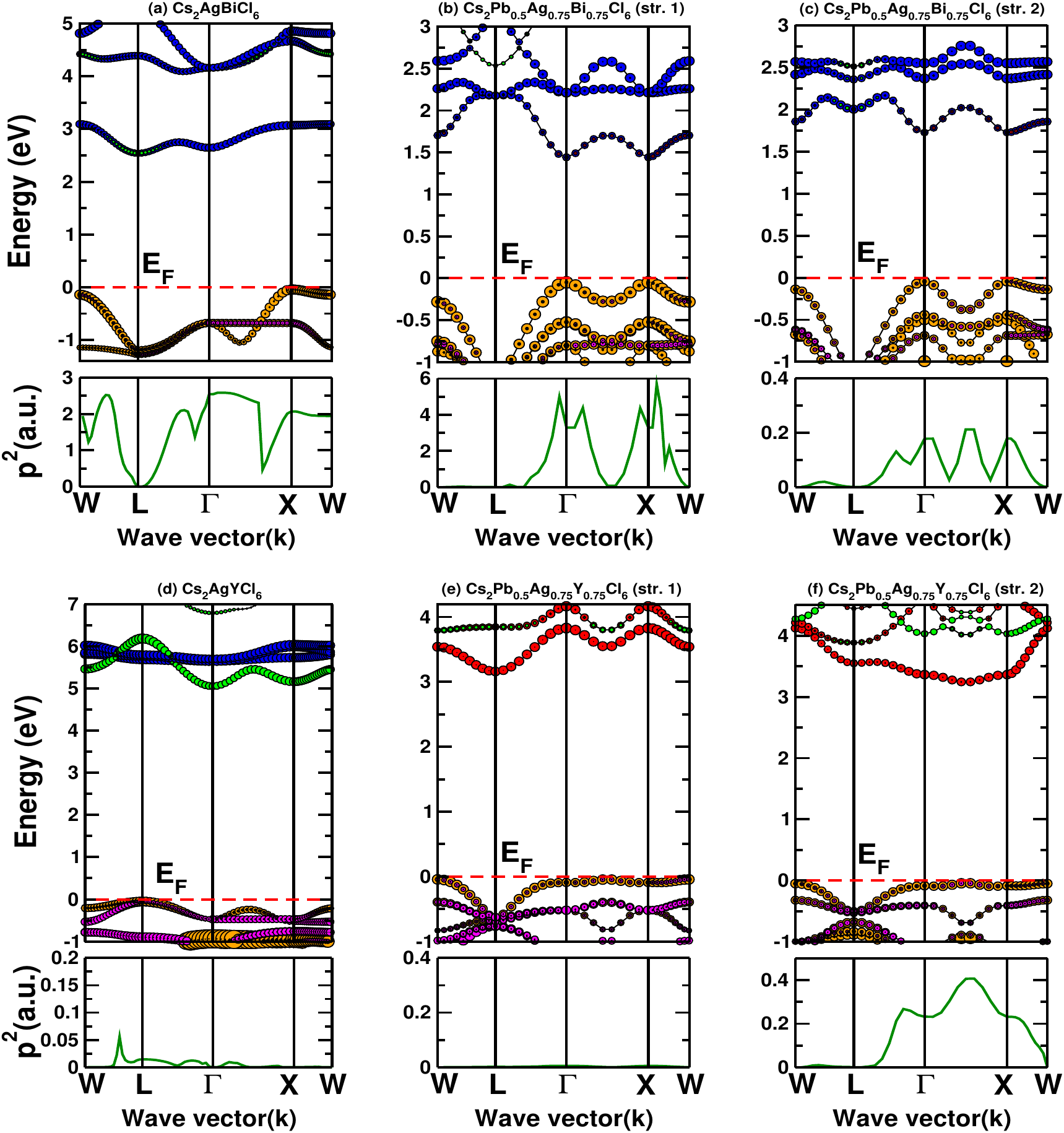}
		\caption{Band structure (above) \& transition probability from VBM to CBM (square of dipole transition matrix elements) (below) for (a) pure Cs$_2$AgBiCl$_6$ and Cs$_2$Pb$_{0.5}$Ag$_{0.75}$Bi$_{0.75}$Cl$_6$ in (b) Structure `1' \& (c) Structure `2' respectively. (d), (e) and (f) shows similar results but for Cs$_2$AgYCl$_6$. All the calculations are done using PBE exchange-correlation functional including spin-orbit coupling (soc) effect. In these plots band gaps are rigid shifted to HSE06+soc calculated values, for convenience. All the energies are plotted taking valence band maxima as zero. Here Cl-p, Ag-d,  B$^{'}$-p (when  B$^{'}$=Bi) or B$^{'}$-d (when B$^{'}$=Y), Ag-s, and Pb-p states are shown in orange, magenta, blue, green, and red coloured circles respectively}
		\label{fig:test2}
	\end{figure*}

	\begin{figure*}[t!]
		\centering
		\includegraphics[width=1\linewidth]{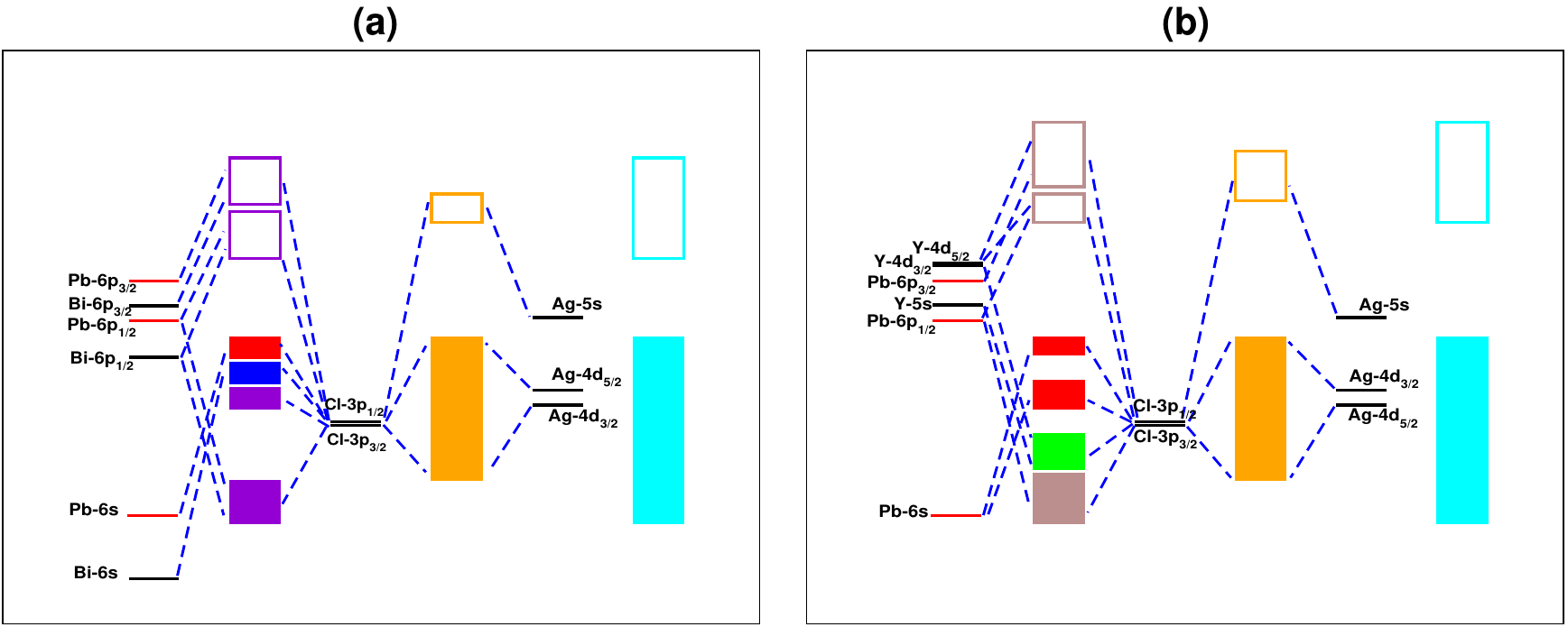}
		\caption{Molecular orbital diagram for Pb doped (a) Cs$_2$AgBiCl$_6$ \& (b) Cs$_2$AgYCl$_6$. Atomic orbitals for Ag, Bi, Y, Cl are marked with black lines and for Pb they are red. Empty boxes signify unfilled conduction bands, whereas filled ones represents valence bands. In figure (a) red coloured bands signify contribution from Pb and Cl only. Blue signifies Bi-Cl interaction, and mixed Bi-Cl, Pb-Cl bands are marked in violet colour. In figure (b) red coloured bands signify contribution from Pb and Cl only. Green signifies Y-Cl interaction, and mixed Y-Cl, Pb-Cl bands are shown in brown colour. Ag-Cl bands are marked in orange colour in both of the figures. Atomic orbital energies are taken from (www.nist.gov). Molecular orbitals are drawn by looking at the orbital projected density of states for these systems (can be found in supplementary)\cite{supplement} }
		\label{fig:test2}
	\end{figure*}

	For a material, chemical stability is a necessary criteria to be satisfied. Here, to determine the chemical stability we calculate the decomposition enthalpy of the compounds in a possible pathway, namely decomposing to the binary halides of the cations, A$_2$PbBB$^{'}$X$_6$ $\rightarrow$ AX + PbX$_2$ + BX + B$^{'}$X$_3$. Details of the quantitative procedure is discussed in the supplementary material (section S1).\cite{supplement} Reverse of this can be seen as the formation energy($\Delta$E$_F$) of the compound from its binary halide salts. We have tabulated the simulated data for all the selected compounds, forming in two different structures in Table 1 \& 2 respectively of the supplementary material.\cite{supplement} Here negative value indicates the stability of the compound. It is noticeable that for a specific compound, energetics of structure `1' and structure `2' are very close (within 1 meV/atom), pointing towards almost equal probability of formation of the concerned compound in any of these two structures at 25\% Pb substitution. This also indicates that disorder can be a common phenomena in these compounds.
	
	\section*{III. Electronic structure}
	
	The double perovskites, reported in the literature till date, face two issues. First, some class have indirect band gap (Cs$_2$AgBiX$_6$ being one of them) and the direct gap is at a much higher energy level than the indirect gap. This results in relatively poor absorption (as phonons are required to assist in absorption). Secondly, they suffer from higher possibility of non-radiative recombination loss. These negative facts restrict them from using as single junction solar cell. By doing a detailed analysis Savory et al. reported that the fundamental mismatch between Ag-d and Bi-s orbitals is the main reason behind this.\cite{savory2016can} They proposed the idea of orbital matching by replacing Ag$^{+1}$ with Tl$^{+1}$ or In$^{+1}$, which have valence ns$^2$ electrons in their cationic state, and showed the band gap to become direct. But as discussed earlier these compounds came out to be thermodynamically unstable.\cite{xiao2017intrinsic}  Later, few compounds of same class (e.g. Cs$_2$AgInCl$_6$) were reported to show direct nature of band gap.\cite{volonakis2017cs2inagcl6,jain2017high} But, the problem with these compounds was their much large optical band gap. The reason behind this is the presence of inversion symmetry in double perovskite structure, which can result in the bands at VBM and CBM to have same parity, thus making the transition from VBM to CBM optically forbidden.\cite{meng2017parity} Here we choose Pb, which is known to have very stable  +2 oxidation state, and ns$^2$ (6s$^2$) valence configuration at that cationic state, similar to Bi. Instead of substituting fully (which practically will result in perovskite CsPbI$_3$), we substitute partially at both B and B$^{'}$ sites.

	We have calculated the electronic properties (band structure, density of states(DOS) etc.) for all the pure and Pb-substituted compounds using PBE exchange correlation functional. We assume that for our systems, shape of the band structure calculated using PBE does not change much compared to others, calculated using more accurate and expensive exchange-correlation functionals.  This is usually the case for most of the materials.\cite{janotti2010hybrid, stroppa2009unraveling} Next, we study the optical properties, in which the use of accurate value of the band gap is extremely important. PBE exchange correlation functional is known to underestimate the band gap value for semiconductors. On the other hand, Heyd-Scuseria-Ernzerhof (HSE) functional is seen to predict the band gap quite close to the experiment.  Spin-orbit coupling (soc) has great effect on the electronic properties of these class of systems. We have included soc in all our calculations. As such, primarily we  did the electronic structure calculation using PBE+soc, whereas, accurate value of the band gap is calculated using HSE06+soc exchange correlation functional (validation and more details of the computational procedure can be found in the supplementary).\cite{supplement} PBE+soc band structure with the relatively more accurate band gap obtained using HSE06+soc exchange correlation functional, is then used for further optical calculations.

	\begin{table*}[t!]
		\begin{ruledtabular}
			\begin{centering} 
				\begin{tabular}{c c c c c c }
					
					\hline
					Compound & $a$ & $b=c$ & $\Delta E_F$ & $E_g$(eV) & SLME  \tabularnewline
					\vspace{0.1 in}  
					& (\r{A}) & (\r{A}) & (meV/atom) & HSE06+soc & ($\eta\%$)   \tabularnewline
					\hline 
					\vspace{0.1 in}  
					\textbf{Structure `1' (cubic)} &  &  &  &  &   \tabularnewline
					\vspace{0.1 in}  
					Cs$_2$Pb$_{0.5}$Ag$_{0.75}$Bi$_{0.75}$Cl$_6$ & 11.11 &  & -96.26 & 1.49(D) & 16.31  \tabularnewline
					\vspace{0.1 in}  
					Cs$_2$Pb$_{0.5}$Ag$_{0.75}$Sb$_{0.75}$Cl$_6$ & 11.02 &  & -76.08 & 1.43(D) & 18.55  \tabularnewline
					\vspace{0.1 in}  
					Cs$_2$Pb$_{0.5}$Au$_{0.75}$Bi$_{0.75}$Cl$_6$ & 11.09 &  & -35.52 & 0.95(D) & 23.80 \tabularnewline\vspace{0.1 in}  
					Cs$_2$Pb$_{0.5}$Au$_{0.75}$Sb$_{0.75}$Cl$_6$ & 11.01 &  & -18.18 & 0.94(D) & 23.64  \tabularnewline\vspace{0.1 in}  
					Cs$_2$Pb$_{0.5}$Au$_{0.75}$Sc$_{0.75}$Cl$_6$ & 10.85 &  & -19.81 & 2.42(D) & 12.38  \tabularnewline\vspace{0.1 in}  
					Cs$_2$Pb$_{0.5}$Au$_{0.75}$Y$_{0.75}$Cl$_6$ & 11.03 &  & -8.15 & 2.45(D) & 12.33  \tabularnewline\vspace{0.1 in}  
					Cs$_2$Pb$_{0.5}$Cu$_{0.75}$Bi$_{0.75}$Cl$_6$ & 10.89 &  & -61.82  & 1.03(D) & 26.84  \tabularnewline\vspace{0.1 in}  
					Cs$_2$Pb$_{0.5}$Cu$_{0.75}$Sb$_{0.75}$Cl$_6$ & 10.82 &  & -41.72 & 1.09(D) & 25.08  \tabularnewline\vspace{0.1 in}  
					Cs$_2$Pb$_{0.5}$Ag$_{0.75}$Bi$_{0.75}$Br$_6$ & 11.64 &  & -62.02 & 1.02(D) & 21.01  \tabularnewline\vspace{0.1 in}  
					Cs$_2$Pb$_{0.5}$Au$_{0.75}$Bi$_{0.75}$Br$_6$ & 11.61 &  & -11.59 & 0.54(D) & 15.60 \tabularnewline\vspace{0.1 in}  
					Cs$_2$Pb$_{0.5}$Ag$_{0.75}$Bi$_{0.75}$I$_6$ & 12.40 &   & -2.46 & 0.50(D) &  16.30 \tabularnewline\vspace{0.1 in}    
					\textbf{Structure `2' (tetragonal)} &  &  &  &  &   \tabularnewline
					\vspace{0.12 in}  
					Cs$_2$Pb$_{0.5}$Ag$_{0.75}$Bi$_{0.75}$Cl$_6$ & 11.24 & 11.05  & -96.60 & 1.77(D) & 12.30  \tabularnewline
					\vspace{0.1 in}  
					Cs$_2$Pb$_{0.5}$Ag$_{0.75}$Sb$_{0.75}$Cl$_6$ & 11.21 & 10.96 & -75.76 & 1.78(D) & 14.07  \tabularnewline\vspace{0.1 in}  
					Cs$_2$Pb$_{0.5}$Au$_{0.75}$Sc$_{0.75}$Cl$_6$ & 11.39 & 10.69 & -18.53 & 2.28(D) & 12.09  \tabularnewline\vspace{0.1 in}  
					Cs$_2$Pb$_{0.5}$Au$_{0.75}$Y$_{0.75}$Cl$_6$ & 11.52 & 10.88 & -9.67 & 2.21(D) & 13.05  \tabularnewline\
					\vspace{0.1 in}  
					Cs$_2$Pb$_{0.5}$Cu$_{0.75}$Bi$_{0.75}$Cl$_6$ & 11.25 & 10.77 & -62.05  & 1.13(D) & 20.40  \tabularnewline\vspace{0.1 in}  
					Cs$_2$Pb$_{0.5}$Cu$_{0.75}$Sb$_{0.75}$Cl$_6$ & 11.24 & 10.68  & -41.82 & 1.25(D) & 16.62  \tabularnewline\vspace{0.1 in}  
					Cs$_2$Pb$_{0.5}$Cu$_{0.75}$Sc$_{0.75}$Cl$_6$ & 11.28 & 10.48  & -39.92 & 2.32(D) & 9.27  \tabularnewline\vspace{0.1 in}  
					Cs$_2$Pb$_{0.5}$Cu$_{0.75}$Y$_{0.75}$Cl$_6$ & 11.24 & 10.68  & -41.82 & 2.25(D) & 11.25  \tabularnewline\vspace{0.1 in}  
					Cs$_2$Pb$_{0.5}$Ag$_{0.75}$Bi$_{0.75}$Br$_6$ & 11.74 & 11.58 & -58.80 & 1.26(D) & 18.52  \tabularnewline\vspace{0.1 in}  
					Cs$_2$Pb$_{0.5}$Ag$_{0.75}$Bi$_{0.75}$I$_6$ & 12.50 & 12.35  & -2.62 & 0.74(D) &  20.12 \tabularnewline\vspace{0.1 in}      \\\hline  
				\end{tabular}
				\par\end{centering}
			\caption{Relaxed lattice parameters ($a$, $b$, $c$), simulated chemical formation energy ($\Delta E_F$), Band gap ($E_g$) and Spectroscopic Limited Maximum Efficiency (SLME) calculated using the hybrid HSE-06 exchange-correlation functional including the effect of spin-orbit coupling (soc) for Pb substituted halide double perovskites. These are the selected list of compounds which are predicted to be most promising out of a more exhaustive list shown in the supplementary material.\cite{supplement} Negative formation energy supports chemical stability of the system.  'D' within the parenthesis indicate the direct nature of band gap. SLME ($\eta\%$) is simulated at room temperature (T=298 K), considering a film thickness of 2$\mu$m. }
			\label{final_data}
		\end{ruledtabular}	
	\end{table*}

	\begin{figure*}[t!]
		\centering
		\includegraphics[width=0.90\linewidth]{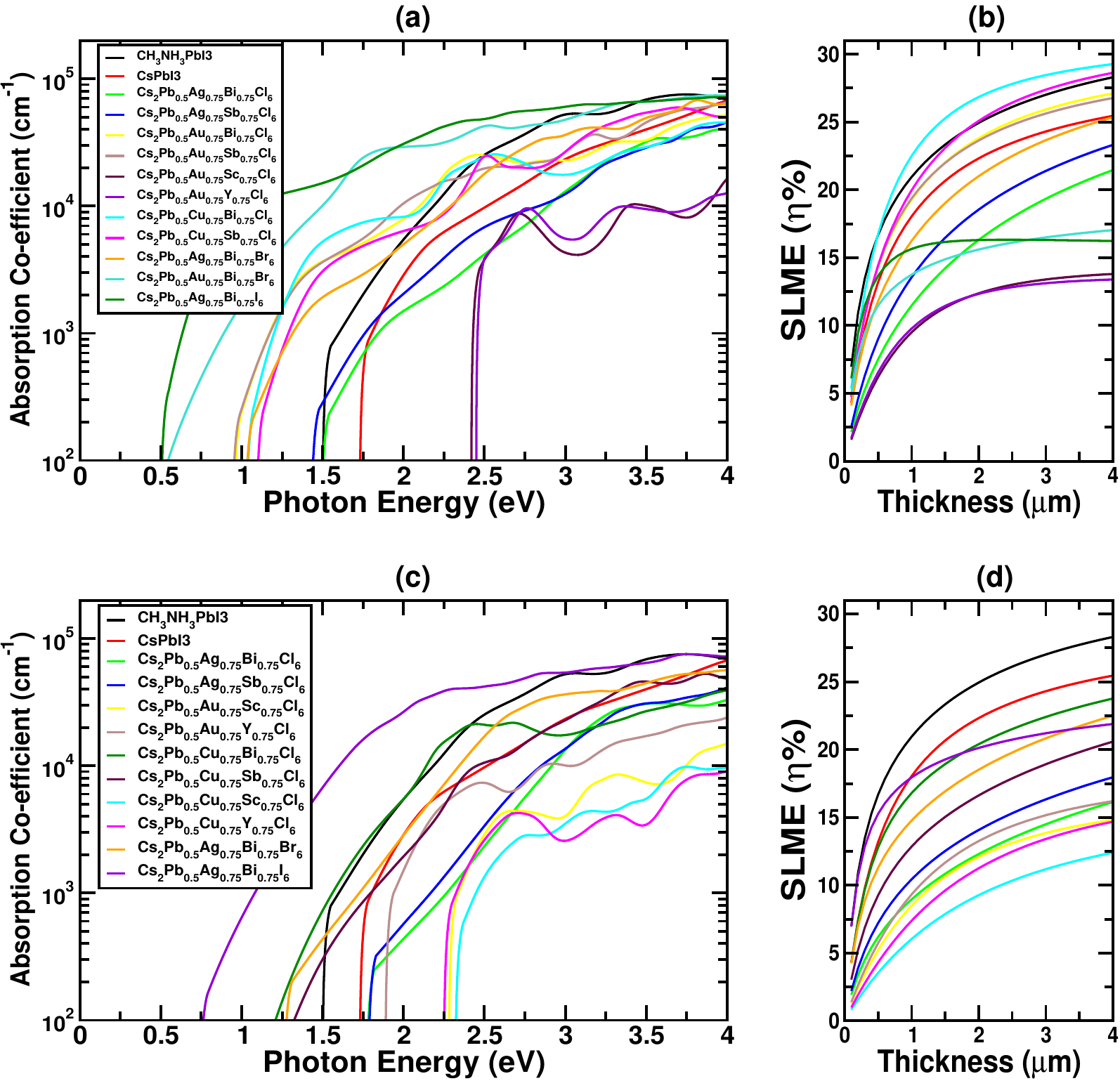}
		\caption{(a) \& (c) Absorption coefficient vs. incident photon energy and (b) \& (d) 'Spectroscopic limited maximum efficiency' vs. film thickness at $298$ K for various compounds (as in Table II) in structure `1' \& `2' respectively. For comparison,  simulated results for state of the art, CH$_3$NH$_3$PbI$_3$ and its inorganic counterpart CsPbI$_3$ is also given.}
		\label{fig:test2}
	\end{figure*}

	We choose to screen a large number of parent compounds (Cs$_2$BB$^{'}$X$_6$, X=Cl, Br, I), with a permutation of group (G)-XI elements (Cu, Ag, Au) at B-site and  G-XV(Sb,Bi) \& G-III(Sc,Y) elements at B$^{'}$-site and hence examine the effect of Pb-substitution. All these compounds show indirect band gap in the parent state. With small Pb substitution most of them transform to direct band gap (See Table 1 \& 2 of supplementary).\cite{supplement} In order to get a deeper insight, we have calculated orbital projected density of states (DOS) for the pure and Pb-substituted systems. We particularly choose two systems namely Cs$_2$AgBiCl$_6$ \& Cs$_2$AgYCl$_6$ and their Pb-substituted counterpart to analyze their electronic properties, here in the manuscript itself. These two systems can be seen as representatives to their respective classes. We have provided all the electronic structure data (band structure, orbital projected density of states, etc.) for the rest of the possible compounds in  supplementary material (see section S3).\cite{supplement}

	Pure Cs$_2$AgBiCl$_6$ is an indirect band gap material having valence band maxima (VBM) \& conduction band minima (CBM) at `X' \& `L' high-symmetry points in reciprocal space as shown in Fig. 2(a). Detailed analysis of orbital projected DOS (shown in  supplement\cite{supplement}) shows that it has Ag-4d/Cl-3p, Bi-6s/Cl-3p character in the VBM and Ag-5s/Cl-3p, Bi-6p/Cl-3p anti-bonding character in CBM. It has been reported that Ag-4d and Bi-6s interaction pushes the valence band to higher energy at `X' point leading to indirect nature of the band gap.\cite{savory2016can} With Pb substitution two different scenario arises for structure `1' and structure `2'. Using the orbital projected density of states, we tried to draw a qualitative molecular orbital diagram for Cs$_2$Pb$_{0.5}$Ag$_{0.75}$Bi$_{0.75}$Cl$_6$ as shown in Fig. 3(a). `Pb' having higher energy 6s atomic orbital, contributes to the valence band pushing it higher than in the parent compound, at $\Gamma$- and X-points. It can be clearly seen from the orbital projected DOS that VBM is mainly composed of Ag-4d/Cl-3p \& Pb-6s/Cl-3p anti-bonding orbitals. Spin-orbit coupling plays an integral role in the electronic structure of these systems, specially for structure `2'. Inclusion of spin-orbit coupling splits the Bi and Pb p orbitals, which results in lowering of the conduction band mainly at `$\Gamma$' and `X'-point, as can be seen from Fig. 2. This causes both structure `1' and structure `2' to have a  direct nature of band gap. Band structure and DOS for other Pb-substituted compounds in this family i.e. Cs$_2$(Cu, Ag, Au)(Sb, Bi)X$_6$ are shown in the supplementary material.\cite{supplement} In case of Sb at B$^{'}$ site, band gap remains similar, only the conduction band width decreases which can be attributed to
	less spread of Sb-p$_{3/2}$ \& Sb-p$_{1/2}$ orbitals compared to Bi-p orbitals. In the compounds with Cu at B site the band gap reduces significantly than its Ag counterpart which can be attributed to Cu-d orbitals having higher energy than Ag-d orbitals, which pushes the valence band even higher. Au having significantly higher electronegativity (2.64) than both Ag (1.93) and Cu (1.9), have the Au-s atomic orbital lower in energy than the other two. This affects in lowering the conduction band energy at 'L' point, making it indirect for both structure `1' and structure `2'. Effect of spin-orbit coupling changes the nature of band gap from indirect to direct for Structure `1' but structure `2' remains indirect in nature. In all the cases, changing the halide from Cl $\rightarrow$ Br $\rightarrow$ I, lifts the VBM as atomic p-orbital energy increases with decreasing electronegativity (with increasing atomic number in halides). Valence band width decreases as the difference in energy between B-d \& X-p orbitals decreases. Due to the more delocalization of p-orbitals going from Cl $\rightarrow$ Br $\rightarrow$ I, bottom of the conduction band widens. These changes results in a lower band gap from chlorides to iodides.

	Cs$_2$AgYCl$_6$ belong to another class. The band structure for this representative compound  is shown in Fig. 2(d). For this compound VBM is at `L' point and mainly composed of Ag-4d/Cl-3p anti-bonding character. CBM mainly comprises of Ag-5s/Cl-3p orbitals at `$\Gamma$'. It also has Y-4d/Cl-3p anti-bonding character. The bonding characteristics for the same can be seen 2-3 eV lower than the VBM. Pb substitution introduces Pb-6s states in the valence band pushing it to higher energy in the middle of `$\Gamma$' and `X', causing a reduction in band gap. Here for structure `1', CBM is at `L' point resulting in indirect nature of band gap. But for structure `2', next nearest Pb-6p,Pb-6p/Cl-3p interaction is seen to drag down the CBM in the middle of $\Gamma$-X line, making the band gap direct.  Band structure and DOS for other Pb-substituted compounds in this family i.e. Cs$_2$(Cu, Ag, Au)(Sc, Y)X$_6$ are shown in the supplementary material.\cite{supplement} Sc having higher electronegativity than Y has Sc-3d atomic orbitals lower in energy than Y. But this energy difference is very small leading to almost similar band gaps. In case of Cu, band pattern remains similar to Ag, giving direct band gap for  structure `2', but reduces the gap significantly. Structure `1' becomes direct in case of Au incorporation at B site, showing a direct gap in the middle of `$\Gamma$' \& `X' and 'W' point. This again can be attributed to much higher electronegativity of Au leading to a significantly lower Au-s atomic orbital than both Ag and Cu. Au in structure '2' is seen to have direct gap similar to Ag and Cu. For chlorides the band gap for these class of compounds remain in the higher visible side. But changing the halide from Cl $\rightarrow$ Br $\rightarrow$ I reduces the gap, putting it well in the visible region. 
	
	For comparison sake, simulated data for band gap (magnitude and nature) with different exchange correlation functionals  and decomposition enthalpies for all the systems are reported in Table 1 and 2 of the supplementary material.\cite{supplement} It can be seen that Pb substitution decreases the band gap considerably for a number of these systems, taking it well into the visible region. However, there are still few systems which are either indirect or unstable (+ve formation enthalpy). At this stage, we shortlist only those compounds which has direct band gap and are stable, to perform further investigation of optical absorption and device efficiency. The list of these selected compounds (in both structure `1' \& `2') are shown in Table II. 
	
	\section*{IV. Optical absorption coefficient \& Spectroscopic Limited Maximum Efficiency (SLME)}
	
	Hybrid halide perovskites has taken the solar cell development to a new level by showing rapid growth in efficiency. This is attributed to its excellent absorption ability and long carrier diffusion length. Careful analysis of electronic structure reveals the reason for strong absorption to be the direct p-p transition from valence band (consisting halide-p orbital) to conduction band (consisting Pb-p).\cite{yin2015superior} The compounds we have discussed so far also show direct p-p transition form VBM to CBM. This motivates us to proceed further and study the necessary optical properties of these systems to be considered as solar absorbers. Here we choose a parameter known as 'Spectroscopic limited maximum efficiency' (SLME) introduced by Yu. et al. \cite{yu2012identification} as a screening parameter. SLME gives the maximum possible solar power conversion efficiency incorporating the nature of the band gap and absorption coefficient for a particular compound. This can be seen as necessary improvement upon the Shockley-Queisser (S-Q) efficiency parameter (more details about SLME and related parameters can be found in the supplementary material).\cite{supplement}
	
	As discussed above, SLME includes two material specific properties. First is the nature of the band gap. For various double perovskites, even after acquiring a direct electronic band gap, it is seen that the optical transition from VBM to CBM was forbidden. This arises because VBM and CBM turn out to have same parity, due to the inversion symmetry present in these structure. In order to check the possibility of optical transition from VBM to CBM for our systems, we have calculated the dipole transition matrix elements, square of which can essentially be seen as the probability of transition. We have plotted the square of the dipole transition matrix elements (p$^2$) for all the systems (below their respective band structure plots) as shown in Fig. 2 of manuscript and section S3 of supplementary material.\cite{supplement} It is clear that the transition from VBM to CBM at direct electronic band gap is allowed (for most of the systems), encouraging us to proceed further with the calculation of next relevant properties, namely the absorption coefficient($\alpha$) and SLME. 
	
	\begin{figure*}[t!]
		\centering
		\includegraphics[width=0.85\linewidth]{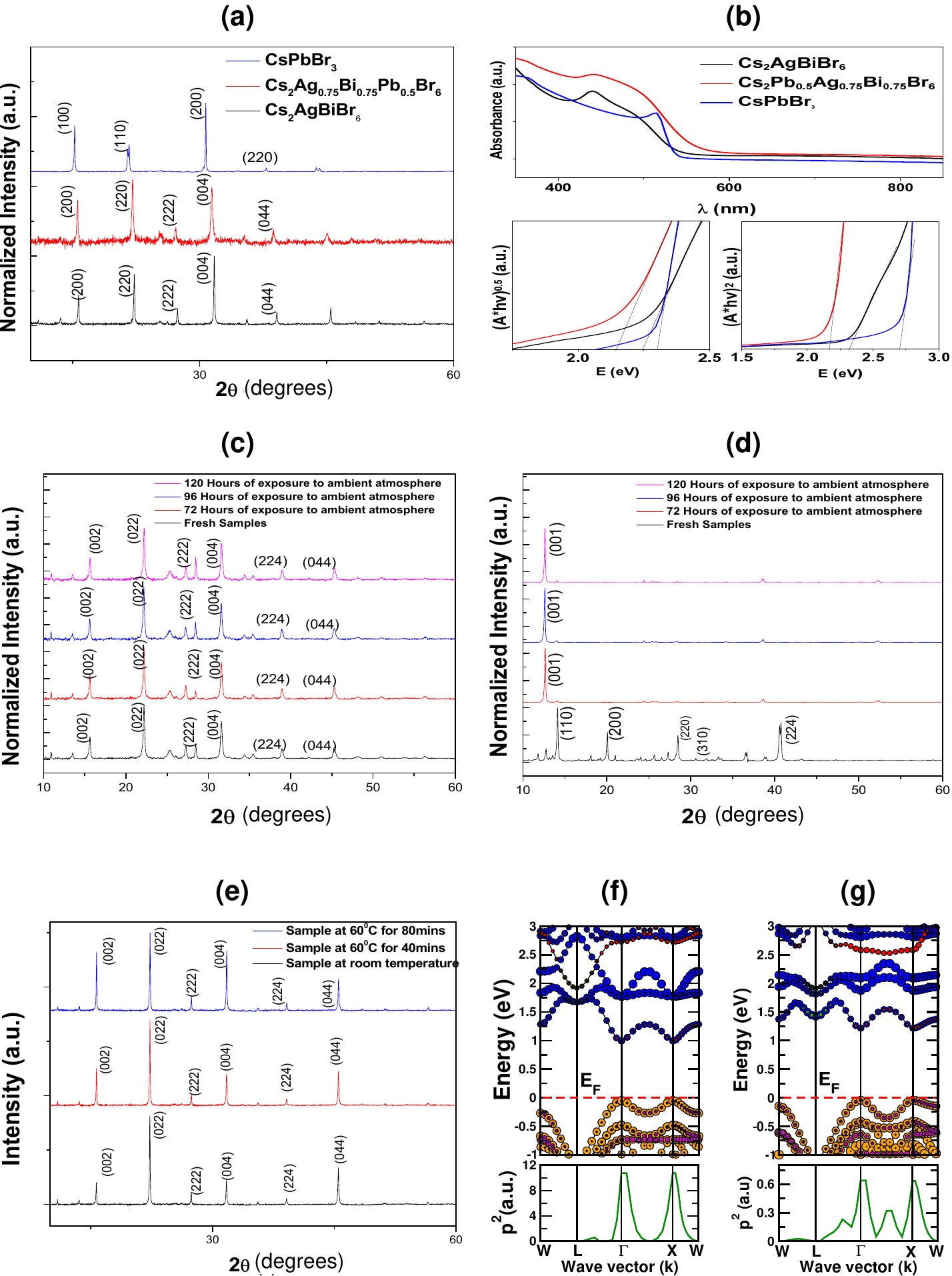}
		\caption{(a) Stacked XRD pattern and (b) UV-visible plot for CsPbBr$_3$,  Cs$_2$Pb$_{0.5}$Ag$_{0.75}$Bi$_{0.75}$Br$_6$,  \&  Cs$_2$AgBiBr$_6$. Figure 5(b) also includes the Tauc plots for the aforementioned compounds. Stacked XRD pattern for (c)
			Cs$_2$Pb$_{0.5}$Ag$_{0.75}$Bi$_{0.75}$Br$_6$ \& (d) CH$_3$NH$_3$PbI$_3$ showing behaviour in humid atmospheric conditions,  and (e) Stacked XRD pattern showing thermal stability for Cs$_2$Pb$_{0.5}$Ag$_{0.75}$Bi$_{0.75}$Br$_6$. Band structure (band gaps scissor shifted to HSE06+soc values ) (above) \& transition probability from VBM to CBM (square of dipole transition matrix elements) (below) for Cs$_2$Pb$_{0.5}$Ag$_{0.75}$Bi$_{0.75}$Br$_6$ in (f) Structure `1' \& (g) Structure `2'.   }
		\label{fig:test2}
	\end{figure*}		
	
	Absorption coefficient ($\alpha$) for a material can be seen as a quantifiable  parameter which basically shows how deep a photon with a particular wavelength can penetrate into the material before it is absorbed. For semiconductors, as the photons below the energy of its band gap takes almost no part in absorption, a sharp rise in the absorption coefficient is expected after a threshold incident photon energy close to its band gap. For a semiconducting material the absorption coefficient($\alpha$) is directly related to its dielectric function by the following equation,
	\begin{equation}
	\alpha(E)=\frac{2\omega}{c}\sqrt{\frac{\sqrt{\epsilon_{re}^2+\epsilon_{im}{^2}}-\epsilon_{re}}{2}}
	\end{equation}	
	Here $E$ is the incident photon energy, $\omega$ is the angular frequency related to $E$, by $E=\hbar\omega$, $\hbar$ is the reduced Planck's constant, $c$ is the speed of light in vacuum. $\epsilon_{re}$ and $\epsilon_{im}$ are the real and imaginary part of dielectric function at that energy. Equation (1) is a simplified version, showing $\alpha$ as scalar. Actually $\alpha$ and the related $\epsilon$'s are second rank tensors. Here we used density functional theory under independent particle approximation to calculate the frequency (energy) dependent dielectric function. Other necessary details and validation of the calculation can be found in supplementary material.\cite{supplement}

	Optical absorption coefficient for all the selected systems (as in Table II) are plotted in Fig. 4($a$)(for structure `1') \& 4($c$) (for structure `2'). For comparison sake we have also plotted the absorption coefficient for state of the art CH$_3$NH$_3$PbI$_3$ and its inorganic counterpart CsPbI$_3$. From the figures it can be seen that, some of the systems have absorption coefficient ($\alpha$) comparable to CH$_3$NH$_3$PbI$_3$. Also, few systems with band gap lower than CH$_3$NH$_3$PbI$_3$ will be able to cover the solar spectrum near the reddish side. Next, we calculated the thickness dependent SLME for all these systems. Figure 4($b$) \& 4($d$) shows the SLME vs. film thickness for our selected systems in two possible structures respectively. Here also, we have plotted similar data for CH$_3$NH$_3$PbI$_3$ and CsPbI$_3$, to see how our systems trade with the state of the art scenario.  Interestingly, one can see that few of our proposed material has SLME comparable to the state of the art CH$_3$NH$_3$PbI$_3$. Even more remarkable are few systems (Cu-based) with even higher efficiency than CH$_3$NH$_3$PbI$_3$. These excellent SLME can essentially be attributed to very high absorption coefficient and having band gaps well in the visible region and very near to ideal S-Q region.
	
	For completeness, we have also tabulated room temperature (298 K) simulated value of few important device  parameters, such as short-circuit current density(J$_{sc}$), open-circuit voltage(V$_{oc}$), current density(J$_{max}$) \& voltage(V$_{max}$) at maximum power output, and fill factor(FF) at film thickness 2$\mu$m in supplement (see Table-3 \& 4).\cite{supplement}

	\section*{V. Experimental Results}
	
	Solution deposition technique	is used to prepare the parent and the Pb-substituted  Cs$_2$AgBiBr$_6$. CsPbBr$_3$ is also prepared for comparative analysis.
	X-Ray Diffraction measurement was carried out to confirm the phase purity and hence stability of these compounds.  Figure 5(a) shows the powder pattern obtained for both the parent compounds CsPbBr$_3$ and Cs$_2$AgBiBr$_6$. Cs$_2$AgBiBr$_6$ crystallizes in 3D double perovskite structure, in cubic Fm-3m space group with unit cell parameter of a = 11.27\r{A}\cite{mcclure2016cs2agbix6} which compares fairly well with the simulated value. Doping of Cs$_2$AgBiBr$_6$ with CsPbBr$_3$ was carried out, by mixing the two solutions in appropriate ratios. The powder diffraction pattern of Pb-doped Cs$_2$AgBiBr$_6$ is also shown in Fig. 5(a) (red curve). As obvious from the powder pattern,  Cs$_{2}$Pb$_{0.5}$Ag$_{0.75}$Bi$_{0.75}$Br$_6$  retain the original double  perovskite structure. 
	
	The band gap of the pristine and the Pb-substituted films were determined using UV-Vis absorption spectroscopy. Figure 5(b) shows the absorption spectra obtained for the pristine and the Pb$^{2+}$ substituted films. The spectra show a steep onset at around 550 nm for Cs$_2$AgBiBr$_6$ and a red shift of 30 nm (absorption onset around 580 nm) was observed for Cs$_2$Pb$_{0.5}$Ag$_{0.75}$Bi$_{0.75}$Br$_6$. From the Tauc plots, direct band gaps of 2.31 eV, 2.13 eV and 2.69 eV was observed for Cs$_2$AgBiBr$_6$,   Cs$_2$Pb$_{0.5}$Ag$_{0.75}$Bi$_{0.75}$Br$_6$ and CsPbBr$_3$ respectively. Their respective indirect band-gaps were 2.21eV, 2.10 eV and 2.27eV. Substitution by Pb$^{2+}$ resulted in a decrease in the band gap of both the parent compounds. It can be observed that, although indirect band gaps for the parent compounds is significantly lower than their respective direct band gaps, the direct and indirect band gap of the Pb doped compound is almost the same. This, in turn, supports our theoretical findings suggesting indirect to direct band gap transition. We have plotted the band structure of Cs$_2$Pb$_{0.5}$Ag$_{0.75}$Bi$_{0.75}$Br$_6$, for both structures '1' and '2' in figure 5(f) and 5(g) respectively, for an one to one comparison with the experimental results. From the band structures it can be clearly seen that, Cs$_2$Pb$_{0.5}$Ag$_{0.75}$Bi$_{0.75}$Br$_6$ in both structures '1' and '2' are direct in nature, confirming an indirect to direct transition with Pb substitution at B and B$^{'}$site of Cs$_2$AgBiBr$_6$. Although, the magnitude of the band gap does not exactly matches with the theoretical values, red shift of the band gap from both Cs$_2$AgBiBr$_6$ and CsPbBr$_3$ is visible here. 
	
	The structural stability of the double perovskite has to be evaluated in order for them to be considered feasible for application in solar cells. In order to study the stability of the material,  Pb$^{2+}$ doped double perovskite films were exposed to ambient atmosphere for a period of 120 hours with humidity of about 55\% on average. For comparison sake we have also checked the stability of CH$_3$NH$_3$PbI$_3$ under exactly the same atmospheric conditions. Powder Diffraction analysis was carried out periodically and the results obtained are shown in the Fig. 5(c) \& 5(d) for Cs$_2$Pb$_{0.5}$Ag$_{0.75}$Bi$_{0.75}$Br$_6$ \& CH$_3$NH$_3$PbI$_3$ respectively. It can be seen that even after 120 hours of full exposure to highly humid conditions the powder patterns remains the same for the former, confirming the robust stability of the material against atmospheric conditions. While CH$_3$NH$_3$PbI$_3$ degrades to binary lead halide after 72 hours, as obvious from the powder diffraction pattern. 
	
	Stability of Pb doped double perovskite films were also checked against extreme thermal condition by exposing the sample to continuous heating at 65$\degree$C for 80 minutes and periodically checking the XRD patterns. The powder patterns at different time interval are shown in Fig. 5(e). The patterns show no significant change in the reflections from the XRD analysis, indicating good thermal stability with respect to temperature. Thus, Cs$_2$Pb$_{0.5}$Ag$_{0.75}$Bi$_{0.75}$Br$_6$ acquire robust atmospheric as well as thermal stability along with promising optical absorption properties, making them a promising candidate for potential solar cell materials, even better than the state of the art CH$_3$NH$_3$PbI$_3$.

	\begin{table*}[t!]
		\begin{ruledtabular}
			\begin{centering} 
				\begin{tabular}{c c c c c c c}
					
					\hline
					Compound & $E_g$(eV) & ($\eta\%$) & ($\eta\%$) & ($\eta\%$) & ($\eta\%$) & $E_g$(eV)   \tabularnewline
					\vspace{0.1 in}  
					& (HSE06+soc) & ($E_g)$  & ($E_g$+10\%) & ($E_g$-10\%) & max. & ($\eta\%$)$_{max}$ \tabularnewline
					\hline 
					\vspace{0.1 in}  
					\textbf{Structure `1' (cubic)} &  &  &  &  &  & \tabularnewline
					\vspace{0.1 in}  
					Cs$_2$Pb$_{0.5}$Ag$_{0.75}$Bi$_{0.75}$Cl$_6$ & 1.49 & 16.31 & 14.53 & 17.82 & 19.05 & 1.10 
					\tabularnewline\vspace{0.1 in}  
					Cs$_2$Pb$_{0.5}$Ag$_{0.75}$Sb$_{0.75}$Cl$_6$ & 1.43 & 18.55  & 16.87 & 19.77 & 20.67 & 1.11
					\tabularnewline\vspace{0.1 in}  
					Cs$_2$Pb$_{0.5}$Au$_{0.75}$Bi$_{0.75}$Cl$_6$ & 0.95 & 23.80  & 24.25 & 22.84 & 24.37 & 1.11   \tabularnewline\vspace{0.1 in}  
					Cs$_2$Pb$_{0.5}$Au$_{0.75}$Sb$_{0.75}$Cl$_6$ & 0.94 & 23.64  & 24.08 & 22.77 & 24.24 & 1.11 \tabularnewline\vspace{0.1 in}  
					Cs$_2$Pb$_{0.5}$Au$_{0.75}$Sc$_{0.75}$Cl$_6$ & 2.42 & 12.38  & 8.30 & 16.69 & 30.51 & 1.16  \tabularnewline\vspace{0.1 in}  
					Cs$_2$Pb$_{0.5}$Au$_{0.75}$Y$_{0.75}$Cl$_6$ & 2.45 & 12.33  & 8.00 & 17.22 & 31.53 &  1.16 \tabularnewline\vspace{0.1 in}  
					Cs$_2$Pb$_{0.5}$Cu$_{0.75}$Bi$_{0.75}$Cl$_6$ & 1.03 & 26.84  & 27.15 & 26.13 & 27.15 & 1.13  \tabularnewline\vspace{0.1 in}  
					Cs$_2$Pb$_{0.5}$Cu$_{0.75}$Sb$_{0.75}$Cl$_6$ & 1.09 & 25.08  & 24.93 & 24.67 & 25.13 & 1.12 \tabularnewline\vspace{0.1 in}  
					Cs$_2$Pb$_{0.5}$Ag$_{0.75}$Bi$_{0.75}$Br$_6$ & 1.02 & 21.01  & 21.07 & 20.61 & 21.08 & 1.11 \tabularnewline\vspace{0.1 in}  
					Cs$_2$Pb$_{0.5}$Au$_{0.75}$Bi$_{0.75}$Br$_6$ & 0.54 & 15.60  & 16.49 & 14.59 & 19.19 & 0.93 \tabularnewline\vspace{0.1 in}  
					Cs$_2$Pb$_{0.5}$Ag$_{0.75}$Bi$_{0.75}$I$_6$ & 0.50 & 16.30  & 18.14 & 14.30 & 29.07 & 1.14  \tabularnewline\vspace{0.1 in}    
					\textbf{Structure `2' (tetragonal)} &  &  &  &  &  &  \tabularnewline
					\vspace{0.1 in}  
					Cs$_2$Pb$_{0.5}$Ag$_{0.75}$Bi$_{0.75}$Cl$_6$ & 1.77 & 12.30 & 9.91 & 14.58 & 18.64 & 1.03
					\tabularnewline\vspace{0.1 in}  
					Cs$_2$Pb$_{0.5}$Ag$_{0.75}$Sb$_{0.75}$Cl$_6$ & 1.78 & 14.07 & 11.47 & 16.54 & 20.74 & 1.11 
					\tabularnewline\vspace{0.1 in}  
					Cs$_2$Pb$_{0.5}$Au$_{0.75}$Sc$_{0.75}$Cl$_6$ & 2.28 & 12.09  & 8.52 & 15.99 & 27.25 & 1.16   \tabularnewline\vspace{0.1 in}  
					Cs$_2$Pb$_{0.5}$Au$_{0.75}$Y$_{0.75}$Cl$_6$ & 2.21 & 13.25  & 9.76 & 17.00 & 27.02 & 1.16  \tabularnewline\vspace{0.1 in}  
					Cs$_2$Pb$_{0.5}$Cu$_{0.75}$Bi$_{0.75}$Cl$_6$ & 1.13 & 20.40  & 19.73 & 20.62 & 20.62 & 1.02  \tabularnewline\vspace{0.1 in}  
					Cs$_2$Pb$_{0.5}$Cu$_{0.75}$Sb$_{0.75}$Cl$_6$ & 1.25 & 16.62  & 15.52 & 17.40 & 17.65 & 0.99  \tabularnewline\vspace{0.1 in}  
					Cs$_2$Pb$_{0.5}$Cu$_{0.75}$Sc$_{0.75}$Cl$_6$ & 2.32 & 9.27  & 6.27 & 12.62 & 23.82 & 1.12  \tabularnewline\vspace{0.1 in}  
					Cs$_2$Pb$_{0.5}$Cu$_{0.75}$Y$_{0.75}$Cl$_6$ & 2.25 & 11.25  & 7.99 & 14.70 & 24.28 & 1.13  \tabularnewline\vspace{0.1 in}  
					Cs$_2$Pb$_{0.5}$Ag$_{0.75}$Bi$_{0.75}$Br$_6$ & 1.26 & 18.52  & 17.47 & 19.26 & 19.37 & 1.03  \tabularnewline\vspace{0.1 in}  
					Cs$_2$Pb$_{0.5}$Ag$_{0.75}$Bi$_{0.75}$I$_6$ & 0.74 & 20.12  & 21.15 & 18.80 & 22.78 & 1.08  \tabularnewline\vspace{0.1 in}     \\  \hline 
				\end{tabular}
				\par\end{centering}
			\caption{Sensitivity of SLME ($\eta\%$) with band gap for Pb substituted double perovskites (structure `1'\& `2'). $E_g$ is the calculated band gap $E_g$ using HSE+soc.. We also tabulated the ideal band gap at which the calculated SLME is maximum, as shown in the last two columns.  SLME ($\eta\%$) is simulated at room temperature (T=298 K), considering a film thickness of 2$\mu$m.}
			\label{final_data}
		\end{ruledtabular}	
	\end{table*}

	\section*{VI. Conclusion and Discussion}
	
	In conclusion, we propose a class of brand new materials which, due to its better efficiency, stability and minimal toxicity, can substitute the state of the art photovoltaic absorber material CH$_3$NH$_3$PbI$_3$ for solar cell applications. These new compounds are double perovskite materials (Cs$_2$BB$^{'}$X$_6$) doped with small amount of Lead (Pb). Pb$^{+2}$ doping at both B and B$^{'}$ sites shows three major +ve effects (i) helps to achieve  indirect-direct band gap transition via orbital matching (ii) pushes the valence band higher, resulting in a significant reduction in band gap and (iii) introduce some new bands at CBM with different parity making the optical transition between VBM to CBM to be allowed. We have also calculated the optical absorption spectra of these systems and found very high absorption coefficient, comparable (sometimes higher) to the state of the art material MAPbI$_3$. Solar efficiency (SLME) calculation on these systems reveal quite high efficiencies for a number of compounds. Calculation of decomposition enthalpy indicates toward stability against possible degradation to binary halides. Apart from simulation, one of the proposed material Cs$_2$Pb$_{0.5}$Ag$_{0.75}$Bi$_{0.75}$Br$_6$ is prepared in our lab and its stability and electronic properties are discussed in detail. Theoretical results matches fairly well with the experimental data, confirming the validity of other theoretically predicted materials. A one-to-one comparison of the measured stability of this particular compound with those of MAPb$_{3}$ shows the former to be much superior : A bottleneck for the later which remains a major concern in the photovoltaic community.
	
	{\it Further ideas to improve the efficiency of  proposed materials :}
	
	Substituting the inorganic Cs element with CH$_3$NH$_3$ or other organic cations may push the efficiency limit even further and make them more stable. Lead halide perovskites have been the centre of attention of the photovoltaic research community for quite a while, with still having issues with toxicity  and stability. Our proposal is a breakthrough to address these concerns and achieve potentially solar efficient materials at 75\% less toxicity and much higher stability. 
	
	It is well known that the solar efficiency of a device is quite sensitive to actual value of the band gap of the material used.  As there are no experimental band gap data available for these newly proposed materials, we have checked the sensitivity of the SLME with its band gap value. (Details of the procedure can be found in the supplementary).\cite{supplement} We have tabulated the SLME for these systems at $\pm$10\% of their HSE06+soc band gap in Table-III. We have also calculated the ideal band gap for each system at which SLME is maximum, as shown in the last two columns.
	Notably,  the ideal band gap falls in the region of 1.0-1.2 eV. In fact, some of the compounds (Cs$_2$Pb$_{0.5}$(Cu,Au)$_{0.75}$(Bi,Sb)$_{0.75}$Cl$_6$, Cs$_2$Pb$_{0.5}$Ag$_{0.75}$Bi$_{0.75}$Br$_6$) shows very high efficiencies, as much as $\sim$31.5\% . From figure 4($a$) \&  4($c$), It can be seen that Cs$_2$Pb$_{0.5}$Ag$_{0.75}$Bi$_{0.75}$I$_6$ has excellent absorption spectra (higher than others), but with low band gap. This significantly reduces the simulated attainable voltage, reducing the efficiency. However, it has a potential to reach very high efficiency, by making the band gap slightly higher (see Table-III). This can be thought to be achieved via lowering the Pb substitution percentage. On the other hand, chloride compounds with G-III (Sc,Y) at B$^{'}$ sites, show very high band gaps (significantly adrift of the ideal region), resulting in poor SLME. Band gap  for such systems reduces with changing the halide from chloride to iodide, which again can  boost the performance. 	
	
	\section*{VII. Methods}
	
	\subsection*{A. Computational methodology:}
	First principles calculations were performed using Density Functional Theory (DFT)\cite{kohn1965self} as implemented in Vienna Ab-initio Simulation Package (VASP).\cite{kresse1996efficiency,kresse1999ultrasoft} Finding the equilibrium structures, calculation of  decomposition enthalpy and other primary electronic structure (band structure, density of states etc.) were done using Perdew-Burke-Ernzerhof (PBE) exchange correlation functional.\cite{perdew1996generalized} Next, the accurate value for the compounds having direct band gap were calculated using Heyd-Scuseria-Ernzerhof (HSE06) functional \cite{krukau2006influence} taking the band edge information from PBE calculated band structures. Spin-orbit coupling has been included in all the calculations. More details and validation of the procedure has been discussed in supplementary.\cite{supplement} Optical absorption coefficients are calculated within the independent particle approximation with PBE exchange correlation functional and then scissor shifted to HSE06 obtained band gap while calculating the SLME. Again more details about procedure and validation can be found in supplementary. 
	
	\subsection*{B. Experimental synthesis:}
	
	All chemicals used were from Sigma Aldrich 99.99\% pure anhydrous grade. A solution in DMSO containing 0.5 mM equivalent of CsBr (106 mg) and 0.25 mM each of AgBr (47 mg) and BiBr$_3$ (112.5 mg) was prepared. This was used as the parent solution for preparation of pure Cs$_2$AgBiBr$_6$ films. Another solution was prepared containing 0.5 mM each of CsBr (106 mg) and and PbBr$_2$ (183.5 mg). This solutions was mixed in appropriate ratios with the parent solution to obtain different percentage of lead doping. Cs$_2$AgBiBr$_6$ and Cs$_2$Pb$_{0.5}$Ag$_{0.75}$Bi$_{0.75}$Br$_6$ films were formed on plain glass slides coated with mesoporous TiO$_2$ following the procedure followed in \cite{greul2017highly}. Mesoporous TiO$_2$ solution was prepared by taking 1:3.5 weight ratio of DyeSol 17.5 NRT TiO$_2$ paste and ethanol, and stirring for around 2 hours. The solution was spin coated on plain glass slides at 3500rpm for 30 seconds and sintered at 500$\degree$C for 15 minutes. The active layers were formed by spin coating warm perovskite solutions (heated at 70$\degree$C) at 200rpm for 25 seconds on the substrates (also heated at 70$\degree$C) and annealing at 270$\degree$C for 10 minutes.In order to synthesize Methyl Ammonium Iodide a solution of 33\% Methylamine in ethanol was taken in a round bottom flask and an equimolar amount of 57\% Hydroiodic acid (HI) was added drop wise to the vigorously stirring solution, maintained at 0$\degree$C, and further stirred for 2 hours.  The solvent was then evaporated under reduced pressure to obtain white crystalline Methylammonium iodide (CH$_3$NH$_3$I) powder. The powder was washed thrice with  Diethyl ether to remove the excess HI and dried in vacuum oven overnight at 50$\degree$C. In order to prepare solution for making CH$_3$NH$_3$PbI$_3$ films, equimolar (1mM each) amounts of CH$_3$NH$_3$I and PbI$_2$ were taken in 1 ml DMSO and spin coated at 3000 rpm for 30 secs on the substrates coated with mesoporous TiO$_2$ and annealed at 100$\degree$C for 10 minutes.  The UV-Visible spectroscopy was done using PerkinElmer LAMBDA 750 to study the optical properties and X-ray Diffraction was done using Philips PW1830 X-Ray Diffraction Spectrometer X-ray Diffractometer (XRD) to study the structural behavior of the materials under exposure to ambient air and temperature over a period of time. \cite{gajdovs2006linear, green2004third, tiedje1984limiting}
	
	\section*{Acknowledgements}
	J.K. acknowledges financial support from IIT Bombay in the form of a Teaching Assistantship. AA  and AYe acknowledges National Center for Photovoltaic Research and Education (NCPRE), IIT Bombay for possible funding to support this research. VS thanks DST-INSPIRE fellowship for the financial support and this research activity is partly supported by Department of Science and Technology, Government of India under the grant number DST/TMD/SERI/S111(G).

\end{document}